\documentclass{pasj00}
\begin{document}
\SetRunningHead{Y. Urata et al.}{Optically dark GRB 051028}
\Received{2006/10/17}%{yyyy/mm/dd}
\Accepted{2007/05/31}%{yyyy/mm/dd}

\title{A multi band study of the optically dark GRB 051028}

%%% Please use the following style in case that sorting by 
%%% affiliation is impossible. 
%
 \author{%
   Yuji \textsc{Urata}\altaffilmark{1},
   Kui-Yun \textsc{Huang}\altaffilmark{2},
   Ping-Hung \textsc{Kuo}\altaffilmark{2},
   Wing-Huen \textsc{Ip}\altaffilmark{2},
   Yulei \textsc{Qiu}\altaffilmark{3},\\
   Keisuke \textsc{Masuno}\altaffilmark{1}, 
   Makoto \textsc{Tashiro}\altaffilmark{1}, 
   Keichi \textsc{Abe}\altaffilmark{1},
   Kaori \textsc{Onda}\altaffilmark{1}, 
   Natsuki \textsc{Kodaka}\altaffilmark{1},  \\ 
   Makoto \textsc{Kuwahara}\altaffilmark{4,5},
   Toru \textsc{Tamagawa}\altaffilmark{5},
   Fumihiko \textsc{Usui}\altaffilmark{6},
   Kunihito \textsc{Ioka}\altaffilmark{7},\\
   Yi-Hsi \textsc{Lee}\altaffilmark{2},
   Jianyan \textsc{Wei}\altaffilmark{3},
   Jinsong \textsc{Deng}\altaffilmark{3},
   Weikang \textsc{Zheng}\altaffilmark{3},
   and
   Kazuo \textsc{Makishima}\altaffilmark{8,4}
}

\altaffiltext{1}{Department of Physics, Saitama University, Shimo-Okubo 255, Sakura, Saitama 338-8570}
\email{urata@crystal.heal.phy.saitama-u.ac.jp}
\altaffiltext{2}{Institute of Astronomy, National Central University, Chung-Li 32054, Taiwan, Republic of China}
\altaffiltext{3}{National Astronomical Observatories, Chinese Academy of Sciences, Beijing 100012, China}
\altaffiltext{4}{Tokyo University of Science, 1-3 Kagurazaka, Shinjyuku, Tokyo} 
\altaffiltext{5}{RIKEN (Institute of Physical and Chemical Research), 2-1 Hirosawa, Wako, Saitama 351-0198} 
\altaffiltext{6}{Japan Aerospace Exploration Agency, Institute of Space and Astronomical Science, Sagamihara, Kanagawa 229-8510} 
\altaffiltext{7}{Department of Physics, Kyoto University, Kitashirakawa, Sakyo-ku, Kyoto 606-8602}
\altaffiltext{8}{Department of Physics, The University of Tokyo, 7-3-1 Hongo, Bunkyo-ku, Tokyo 113-0033} 

%% `\KeyWords{}' always has to be placed before `\maketitle'.
 %Do NOT move this preamble from here!

\maketitle

\begin{abstract}

Observations were made of the optical afterglow of GRB 051028 with the
Lulin observatory's 1.0 m telescope and the WIDGET robotic telescope
system.  R band photometric data points were obtained on 2005 October
28 (UT), or 0.095-0.180 days after the burst.  There is a possible
plateau in the optical light curve around 0.1 days after the burst;
the light curve resembles optically bright afterglows (e.g. GRB
041006, GRB 050319, GRB060605) in shape of the light curve but not in
brightness. The brightness of the GRB 051028 afterglow is 3 magnitudes
fainter than that of one of the dark events, GRB 020124.
Optically dark GRBs have been attributed to dust extinction within the
host galaxy or high redshift.
However, the spectrum analysis of the X-rays implies that there is no
significant absorption by the host galaxy. Furthermore, according to
the model theoretical calculation of the Ly$\alpha$ absorption to find
the limit of GRB 051028's redshift, the expected $R$ band absorption
is not high enough to explain the darkness of the afterglow.  While
the present results disfavor either the high-redshift hypothesis or
the high extinction scenario for optically dark bursts, they are
consistent with the possibility that the brightness of the optical
afterglow, intrinsically dark.

%Please read ``IMPORTANT NOTICE'' carefully before preparing a manuscript. 
\end{abstract}

\section{Introduction}

In recent years, the {\it BeppoSAX} and {\it HETE-2} have provided
quick positional information for a number of gamma-ray bursts (GRBs)
with a typical positional accuracy of $\sim10'$, which has led to
rapid follow-up observations in the optical and near infrared
frequencies.  The {\it Swift} satellite has opened the door to the
making of high sensitivity X-ray afterglow observations with an X-ray
telescope in the early stage of the afterglows. This revealed that 
almost all of the GRBs had an X-ray afterglow: the positions of the
GRBs could be measured within a precision of 10 arcseconds.
Nevertheless, the afterglow associated with about half of the promptly
localized GRBs was either very faint ($>$ 23 mag 1 day after the
burst; Fynbo et al. 2001), or non-existent
%there was no optical afterglows
\citep{hete-2}.  Such events are broadly termed ``optically dark
GRBs''. To be more precise, in this paper we define a GRB to be
``optically dark'' if its optical afterglow is fainter than 23 mag at
1 day after the burst. Typical optically dark GRBs include GRB030115
and GRB021211.  The afterglow of GRB030115 was extremely red; although
a near infrared counterpart with $K\sim$19 was detected 1 day after
the burst, no optical afterglow brighter than 20 mag. was detected,
even at 0.1 days.  In the case of GRB021211, the afterglow showed a
rapid decay until around 0.1 days, fading to $>22$ mag. by the next
day.

The observations of the afterglow of a GRB via X-rays, through radio
frequencies can be interpreted by the fireball models. In general a
shock produced by the interaction of relativistic ejecta with the
circumburst medium will lead to the radiation of broadband synchrotron
emission.  Assuming this widely accepted hypothesis to be true, there
are three possible explanations for optically dark GRBs: (1) They have
such high redshifts ($z>5$) that optical afterglows suffer from strong
Lyman absorption \citep{Heise2}: (2) The optical afterglow has been
extinguished by dust in the vicinity of the GRB or in the star-forming
region in which the GRB occurs (Piro et al. 2002; Klose et al. 2002):
(3)The optical afterglow exhibits rapid decay from a very early phase,
such as has been reported for GRB021211 and GRB020124 (Crew et al
2003, Berger et al 2002, Yamazaki et al 2003).

In this paper we present an analysis of the optical and X-ray
afterglow of an optically dark event, GRB 051028.  At 13:36:01.47 UT
on 28 Oct 2005, the {\it HETE-2} FREGATE and WXM \citep{wxm}
instruments detected a bright GRB \citep{hete}. The burst triggered
the operation of FREGATE in the 30-400 keV energy band. The burst
duration (T90) was 16 seconds in both the 30-400 keV band and the 7-30
keV band. Ground analysis of the WXM data found a 90\% confidence
error region measuring $33'\times18'$ with corners at the following
coordinates:
$\alpha(2000)=01^{\rm h}50^{\rm m}19^{\rm s}.6, \delta(2000)=+47^{\circ}41'06''$,
$\alpha(2000)=01^{\rm h}47^{\rm m}01^{\rm s}.1, \delta(2000)=+47^{\circ}38'02''$,
$\alpha(2000)=01^{\rm h}46^{\rm m}58^{\rm s}.3, \delta(2000)=+47^{\circ}55'55''$,
and 
$\alpha(2000)=01^{\rm h}50^{\rm m}$
17$^{\rm s}$.5, $\delta(2000)=+47^{\circ}58'58''$.
%$\alpha(2000)=01^{\rm h}50^{\rm m}17^{\rm s}.5, \delta(2000)=+47^{\circ}58'58''$ .
%%$\alpha(2000)=01^{\rm h}50^{\rm m}17^{\rm s}.5, \delta(2000)=+47^{\circ}58'58''$.
%
%		
The 30-400 keV fluence of GRB 051028 is $6\times 10^{-6} {\rm
erg/cm^{2}}$ ; the 2-30 keV fluences is $6\times 10^{-7} {\rm
erg/cm^{2}}$. The hardness ratio allows us to classify this burst as a
classical GRB \citep{hete}.
{\it Swift} XRT started to observe the field about 7.1 hours after the
burst and found the X-ray afterglow at $\alpha(2000)=01^{\rm h}48^{\rm
m}15^{\rm s}.1, \delta(2000)=+47^{\circ}45'12''.5$ with an uncertainty
of $6''$ (90\% containment) \citep{swift}. The optical afterglow was
also reported by \citet{ot} at the coordinates of $\alpha(2000)=01^{\rm
h}48^{\rm m}15^{\rm s}.01, \delta(2000)=+47^{\circ}45'09''.2$.

\section{Observations}

Optical observations were made by the East-Asia GRB Follow-up
Observation Network
(EAFON\footnote{http://cosmic.riken.jp/grb/eafon/}; \cite{eafon})
using two kinds of telescopes.

\subsection{Pre-GRB observation with WIDGET}

We observed the error region of GRB 051028 \citep{hete} with the very
wide-field camera WIDGET (Abe et al 2006; Tamagawa et al 2005).
WIDGET is a robotic telescope. It monitors the {\it HETE-2}
field-of-view and can detect GRB optical flashes or possible optical
precursors.  The filed-of-view is $62^{\circ}\times62^{\circ}$, which
covers about 80\% of the {\it HETE2} WXM observing field. The system
consists of a 2k$\times$2k Apogee U10 CCD camera and a Canon EF 24mm
f/1.4 wide-angle lens. WIDGET has been in operation at the Akeno site
(Latitude and Longtitude are $+35~{\circ}47'$ and $138^{\circ}30'$,
respectively) since June 2004.  WIDGET monitored the GRB 051028 region
by repeated unfiltered 5-second exposures taken 16.0 min and 11.2 min
before the burst \citep{widget2}.

\begin{figure}
  \begin{center}
    \FigureFile(80mm,80mm){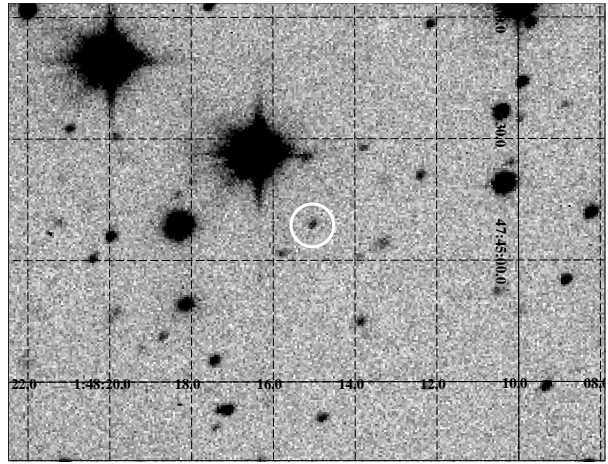}
    %%% \FigureFile(width,height){filename}
  \end{center}
  \caption{$R$ band image of the GRB 051028 field obtained at the Lulin observatory with a 300 s exposure time.
The circle near the center of the image indicates the afterglow.}\label{fig:image}
\end{figure}

\subsection{Follow-up observation at Lulin}

We started optical follow-up observations using Lulin's One-meter
Telescope (LOT) (\cite{lot}) 0.094 days after the burst (i.e., 12 min
after the {\it HETE-2} position alert). This is the fastest follow-up
observation time possible with a meter-sizes-telescope.  This
telescope was installed in September 2002 on the summit of Mount Lulin
($120^{\circ}52'25''$ E, $23^{\circ}28'7''$ N, H$=2862$ m) in central
Taiwan by the Institute of Astronomy of National Central University.
Photometric images were obtained with a PI1300B CCD camera
($1340\times1300$ pixels: $11'.5\times11'.2$ field of view). A samples
is shown in figure 1. To cover the entire {\it HETE-2} WXM error box
($33'\times18'$), we imaged the 8-field mosaic with 300 sec exposures
in the R band.  Due to the darkness and to there being no clear
variability during the early part of our observations as shown in
figure 2, we could not quickly identify the afterglow by analysis at
the observing site.  The obtained R band data is described in Table 1.

\section{Analysis and Results}

\subsection{WIDGET}

The data reduction of the WIDGET images were performed in the standard
manner.  Each WIDGET image taken around the GRB position was compared
with non-saturated bright stars in the Tycho-2.0 position catalog. The
rms deviation around the fit to the positions of 8 reference stars was
$231''$.  We did not find any optical emission from the afterglow
position (Jelinek et al. GCN 4175). The 1-sigma limiting magnitude of
each frame derived from the Tycho-2 catalog was around V=10.3 mag.

\begin{figure}
  \begin{center}
    \FigureFile(80mm,80mm){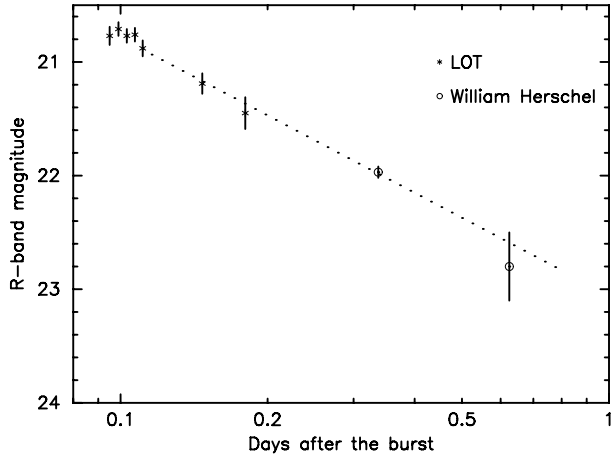}
    %%% \FigureFile(width,height){filename}
  \end{center}
  \caption{$R$ band light curve based on the photometry of the Lulin (LOT) photometry. The dashed line indicates the best fit function for the lightcurve between 0.11 and 0.63 days after ther burst.}\label{fig:rlc}
\end{figure}

\subsection{LOT}

  A standard routine, including bias subtraction, dark subtraction, and
flat-fielding corrections with appropriate calibration data was
employed to process the data using IRAF. 
As shown in figure 1, the afterglow can clearly be seen in the $R$ band
images. Flux calibrations were performed using the APPHOT package in IRAF,
referring to the standard stars suggested by Henden (2005).  For each
data set, the one-dimensional aperture size was set to 4 times as large as
the full-width at half maximum of the objects.  The magnitude of error for
each optical image is estimated as $\sigma_{\rm e}^{2}=\sigma_{\rm
ph}^{2} + \sigma_{\rm sys}^{2}$, where $\sigma_{\rm ph}$ represents the
photometric errors for the GRB051028 afterglow, estimated from the output
of IRAF PHOT, and $\sigma_{\rm sys}$ is the photometric calibration
error estimated by comparing our instrumental magnitudes for the 7
standard stars over the 9 frames.

Figure 2 shows the $R$ band light curve of the GRB 051028 afterglow.
There is a clear plateau phase seen about 0.1 days after the burst.
This early phase plateau is often seen in optically bright afterglows, such as
with GRB 041006 (Urata et al. 2006), GRB 021004 (Urata et al. 2006)
and GRB 050319 (Huang et al 2006). Around 0.11 days after the burst,
the optical afterglow started to decay. We tried to
fit the decaying $R$ band light curves using 
%
%a simple power law of a
%form proportional to $t^{\alpha}$, where t is the time after the onset
%of the burst onset and $\alpha$ is the decay index.  %
%
as a simple power law of the time t after the onset of the burst,
$t^{\alpha}$, where $\alpha$ is the decay index.  We have obtained
$\alpha=-1.06\pm0.04$ with a reduced chi-squared $(\chi^2/\nu)$ of
0.029 for $\nu=$ 1.  In order to better constrain the late-time ($>$
0.3 day) behavior of the light curve, we combined our data with the
two $Rc$-band photometric points taken by the Willam Herschel
telescope; $R$=$21.97\pm$0.05 mag. at 0.337 day, and $R$=22.8$\pm$0.3
mag. at 0.625 day \citep{castro}.  We again successfully fitted the
combined $R$ band light curve with a single power law, for which the
decay index is $-0.904\pm0.037$ with $\chi^{2}/\nu$=0.33 for $\nu=3$.

\begin{table}
  \caption{Lulin photometric result.}\label{tab:first}
  \begin{center}
    \begin{tabular}{ccc}
\hline \hline
Delay (days) & Filter & Magnitude \\ \hline 

0.095  & R & $ 20.77\pm0.08$\\	
0.099  & R & $ 20.71\pm0.06$\\	 
0.103  & R & $ 20.77\pm0.06$\\	
0.107  & R & $ 20.76\pm0.06$\\	 
0.111  & R & $ 20.88\pm0.07$\\	
0.147  & R & $ 21.19\pm0.09$\\	
0.180  & R & $ 21.45\pm0.14$\\	
0.262  & R & $ 21.80\pm0.13$\\ \hline

    \end{tabular}
  \end{center}
\end{table}

\subsection{Swift/XRT}

In order to compare the X-ray afterglow with the optical afterglow, we also
analyzed X-ray data taken by {\it Swift}/XRT.  The data for GRB
051028 were downloaded from the {\it Swift} archive and reduced by
running version 0.10.3 of the xrtpipeline reduction script from the
HEAsoft
6.0.6\footnote{http://heasarc.gsfc.nasa.gov/docs/software/lheasoft/}
software package. However for the four series of observations, the
significance was close to 3 $\sigma$, less than expected from 1
set of XRT data. We then analyzed only the first set of XRT data, for which
observation started at 7.1 hours after the burst.
Spectral response files were generated using the {\tt xrtmkarf} task and the
latest calibration database files (CALDB version 8, 2006-04-27).

The X-ray light curve in the $0.5-5.0$ keV band is a reasonably fit to
a power law model with $\alpha=-1.37\pm0.38$ and $\chi^2/dof=1.00/7$,
which is consistent or slightly steeper than that of the optical
value.  The $0.5-5.0$ keV spectrum is well fitted by an absorbed power
law where the photon index $\Gamma=2.3^{+0.5}_{-0.4}$ with an
absorbing column of $N_{\rm H}=3.1^{+0.2}_{-0.1}\times10^{22} {\rm
cm}^2$ and $\chi^2/dof=0.67/19$.  According to \citet{nh}, the
galactic column density of this line of sight is estimated to be
$1.2\times10^{21} {\rm cm}^2$. 
The mean flux during the observation is
$1.08^{+0.24}_{-0.73}\times10^{-12}$ erg/cm$^2$/s, which is about
1 order fainter than those of optically bright GRB's X-ray afterglows,
such as GRB 050319\citep{x050319}, GRB 051111\citep{x051111} and GRB
060124\citep{x060124}.
These light curve and spectrum analyses indicate that
this X-ray afterglow behavior agrees with the standard model in terms
of the relation between the temporal and spectral indices, assuming
that we are observing a spherical fireball in a frequency range above
that of synchrotron cooling (Sari et al. 1999).
\begin{figure}
  \begin{center}
    \FigureFile(80mm,80mm){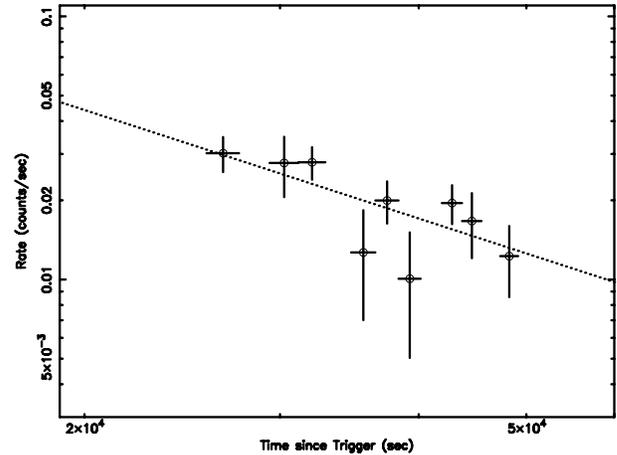}
    %%% \FigureFile(width,height){filename}
  \end{center}
  \caption{X-ray lightcurve taken by {\it Swift}/XRT. The dashed line shows the best fit power law function.}\label{fig:xlc}
\end{figure}

\begin{figure}
  \begin{center}
    \FigureFile(80mm,80mm){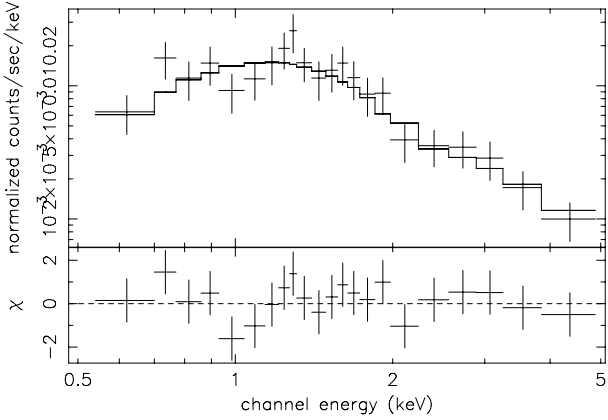}
    %%% \FigureFile(width,height){filename}
  \end{center}
  \caption{X-ray spectrum taken by {\it Swift}/XRT. The spectrum is well fitted by the absorbed power law with photon index $\Gamma = 2.3^{+0.5}_{-0.4}$, absorbing column $N_{H}=3.1^{+0.2}_{-0.1}\times 10^{22}$ cm$^{2}$.}\label{fig:xspe}
\end{figure}

\section{Discussion}

LOT was used to detect the optically dim afterglow of GRB 051028. The
brightness around $0.1$ days after the burst is about 3 magnitudes
fainter than that of the dark GRB 020124 \citep{berger}. The temporal
evolution of the optical afterglow shows a plateau phase around 0.1
days after the burst. These features, with the exception of the
brightness, resemble those of recent optically bright afterglows. The
X-ray afterglow is also 1 order fainter than those of optically bright
GRB's X-ray afterglow. In contrast with one typical, optically dark
event, GRB 021211, the light curve of GRB 051028 shows the usual
pattern of temporal decay, with an index of $\alpha=-0.904$. This is
different from optically dark GRBs, which show rapid decay
from very early phases, such as for GRB 021211.

The optical darkness of the GRB 051028 optical afterglow may
instead be a result of high circumburst extinction.  However the
column density $N_{\rm H}$ shows no significant excess against the
Galactic value. The observed $N_{\rm H}$ (90\% confidence level) is
consistent with that of the Galactic value.  Since this value is
insufficient to explain the optical darkness of dark GRBs, the
extinction model of optically dark GRBs is also not applicable to the
present case. These results are supported the SCUBA observations of
several dark GRBs: the sub-mm results suggest that the optically dark
GRBs do not occur in particularly dusty environments \citep{submm}.

Although the redshift of GRB 051028 was not determined from optical
spectroscopic observation, a value of pseudo-$z =3.7\pm1.8$ can be
derived for this burst using the pseudo-$z$ estimator developed by
Pelangeon et al (2006).  Even assuming the highest allowed redshift
($z=5.5$), the Ly$\alpha$ line and continuum absorption is 
expected to affect the $R$ band flux of the afterglow only by $\sim2$
mag. 
In this calculation, the optical depth is calculated following
\citet{yoshii}, and the spectral index, as computed from the X-ray
afterglow, is fixed at $\beta=-1.3$.
This calculation also successfully explains the drop in the B band in
the spectra of the GRB 050319 ($z=3.24$) afterglow \citep{050319}.
Since the expected $R$ band absorption is not high enough to explain
the darkness of the afterglow, it is inappropriate to use the high-z
scenario for optically dark GRBs, at least for the particular case of
GRB 051028.
The afterglow spectral index $\beta_{ox}$ at 11 hours after the burst
derived from X-ray and optical data is also usefull indicator of the
opticall darkness as Jakobsson et al (2004).  For the current event, the
index $\beta_{ox} \sim -0.6$ agrees with the standard afterglow model
and imply that the optical darkness is unlikely to support high-z and
obscuration.

%Furthermore the afterglow spectral index $\beta_{ox} \sim -0.6$ derived
%from X-ray and optical data agrees with the standard afterglow model
%As Jakobsson et al (2004) claimed, the index supports that the optical
%darkness is unlikely due to high-z and obscuration.

\begin{figure}
  \begin{center} 
\FigureFile(80mm,80mm){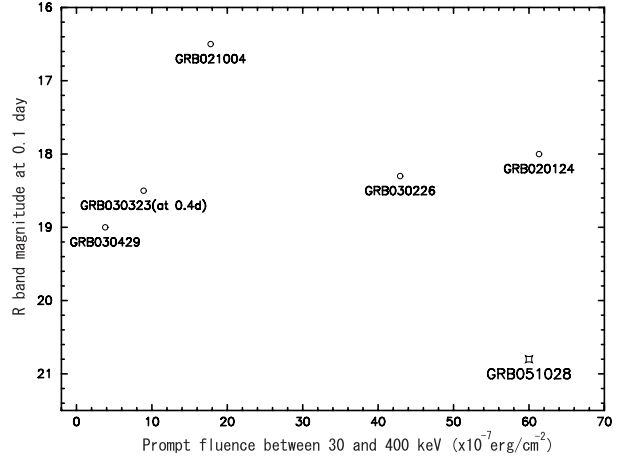} 
%%% \FigureFile(width,height){filename} 
  \end{center} 
\caption{The brightness in the R band at 0.1 days after the burst
plotted against the prompt fluence between 30 and 400 keV. The prompt
fluence come from Hurley et al (2005) for GRB 051028 and Sakamoto et al
(2004) for others}\label{fig:gamma}
\end{figure}

The above discussion suggests that the brightness of the GRB 051028
afterglow is intrinsically optically dark although the prompt fluence
is brighter than that of optically bright events, such as GRB 021004.
The brightness of X-ray afterglow also supports this hypothesis.
Figure 5 shows the R band brightness at 0.1 days against the prompt
fluence.  It can be seen that they have the same redshift range as does
GRB 051028 ($z=3.7\pm1.8$) detected by {\it HETE-2}. This result
implies that the afterglow emission mechanism or the energy conversion
to the afterglows may be the origin of the darkness.
%%%%
{\it Swift} has formulated the canonical X-ray afterglow behavior,
which has led to the observation of early optical afterglows.  These
X-ray and optical light curves show rapid and shallow decay in the
early phase.
%%%%%
These various variablities may be explained by the standard forward
shock scenario, with some additional components, such as continuous
activities related to the central engine, energy injection, patch
shell and 2 jet models (e.g. \citet{ioka}).
%%%
The t$<0.1$ days plateau phase of GRB 051028 could be explained by
energy injection within the context of forward shock model.
%
%As comprehensive discussion of the shallow decay phase with energy
%injection scenario explains the t$<0.1$ days plateau phase of GRB
%051028 as due to energy injection with the standard forward shock
%model.
%%%
In a case of less energy input, there are two expected features: (1)
the brightness of the optical afterglow will be dim, and (2) the
temporal behavior will have a shorter shallow decay phase than those
of bright afterglows, which is similar to the pure standard model.
The shallow decay period of the current GRB 051028 is obviously less
than that of the bright afterglow.  While the bright event of GRB
050319 has a longer shallow decay phase ($\sim 1$days), the afterglow
of GRB 051028 shows the classical temporal decay pattern
($\alpha=-0.9$) from 0.1 days after the burst. 

\section{Conclusion}

We made optical observations using both {\it WIDGET} and the Lulin 1 m
telescope. Based on our optical follow-up observation, it can be seen
that the optical afterglow shows a possible plateau phase at 0.1 days,
which is similar in behavior but not in brightness to optically bright
afterglows (e.g. GRB 041006, Urata et al. 2007; GRB 050319, Huang et
al. 2007; GRB 060605, Deng et al. 2007). The brightness is 3
magnitudes fainter than that of the optically dark GRB 020124. The
X-ray spectrum analysis implies that there is no significant
extinction by the host galaxy.  Furthermore, according to the model
calculation of Ly$\alpha$ absorption limit of GRB 051028's redshift,
the expected $R$ band absorption is not high enough to explain the
darkness of the afterglow.  These arguments indicate that the
faintness of the afterglow of GRB 051028, that is the optically
darkness of the burst, is not due to its being obscured by dust but
because it is intrinsically dim.

\section*{Acknowledgment}
We thank all the staff and observers of the Lulin telescope for
various arrangements in realizing this observation. This work is
supported by NSC 93-2752-M-008-001-PAE and NSC 93-2112-M-008-006. Y.U
acknowledge support from the Japan Society for the Promotion of
Science (JSPS) through JSPS Research Fellowships for Young Scientists.

\end{document}